\documentclass{article}

\usepackage{arxiv}

\usepackage[utf8]{inputenc} 
\usepackage[T1]{fontenc}    
\usepackage{hyperref}       
\usepackage{url}            
\usepackage{booktabs}       
\usepackage{amsfonts}       
\usepackage{nicefrac}       
\usepackage{microtype}      
\usepackage{graphicx}
\usepackage[usenames,dvipsnames]{xcolor}
\usepackage{multirow}
\usepackage{amsmath}
\usepackage{setspace,caption}

\title{Towards a calibration of laboratory setups for grazing incidence and total-reflection X-ray fluorescence analysis}

\author{
  Philipp Hönicke \\
  Physikalisch-Technische Bundesanstalt (PTB)\\
  Abbestr. 2-12\\
  10587 Berlin, Germany \\
  \texttt{philipp.hoenicke@ptb.de} \\
   \And
  Ulrich Waldschläger \\
  Bruker Nano GmbH\\
  Am Studio 2D\\
  12489 Berlin, Germany \\
  \texttt{Ulrich.Waldschlaeger@bruker.com} \\
 \AND
  Thomas Wiesner \\
  Physikalisch-Technische Bundesanstalt (PTB)\\
  Abbestr. 2-12 \\
  10587 Berlin, Germany \\
   \And
  Markus Krämer \\
  AXO DRESDEN GmbH\\
  Gasanstaltstr. 8b \\
  01237 Dresden, Germany \\
 \AND
  Burkhard Beckhoff \\
  Physikalisch-Technische Bundesanstalt (PTB)\\
  Abbestr. 2-12\\
  10587 Berlin, Germany \\
}

\begin{document}
\maketitle

\begin{abstract}
Grazing-Incidence X-ray fluorescence (GIXRF) analysis, which is closely related to total-reflection XRF, is a very powerful technique for the in-depth analysis of many types of technologically relevant samples, e.g. nanoparticle depositions, shallow dopant profiles, thin layered samples or even well-ordered nanostructures. However, the GIXRF based determination of the depth-dependent information about the sample is usually based on a modelling of the experimental data. This requires profound knowledge of the geometrical parameters of the setup employed, especially the incident beam profile as well as the detector aperture parameters. Together they determine the incident angle dependent so-called effective solid angle of detection which must be known in order to model any experimental data set. In this work, we demonstrate how these instrumental parameters, which are typically not known with sufficient accuracy, can be determined using dedicated experiments with a well-known calibration sample. In addition, this paves the way for a full calibration of the setup, as also information about other parameters, e.g. the incident photon flux is gained. Here, we are using a commercially available instrument for this demonstration but the principle can also be applied for other GIXRF setups.

\end{abstract}

\section{Introduction}
In recent years, both the total-reflection X-ray fluorescence (TXRF) \cite{Yoneda_1971,Wobrauschek2007} and the grazing incidence X-ray fluorescence (GIXRF) techniques \cite{K.N.Stoev1999} have been widely used and applied to many different fields of research \cite{Bohlen2009}, including nanoelectronics \cite{V.Soltwisch2018, P.Hoenicke2009,Y.Kayser2015}, nanoparticle characterization \cite{A.Singh2017,vonBohlen2009}, thin layered materials \cite{Hoenicke2019,Maderitsch_2018,Pessoa_2018,H.Rotella2017a,B.Caby2015,M.K.Tiwari2015,D.Ingerle2014a} and even to biological or solid-liquid interface applications \cite{E.Schneck2010,M.Bruecher2010}. Both techniques are based on the interference between incident and reflected photons, resulting in an X-ray standing wave (XSW) field that provides excellent excitation conditions at the reflecting surface with minimal background signals. When tuning the incident angle, the intensity distribution in the XSW changes, which allows depth-dependent information to be derived. Even though TXRF is a well-established technique in laboratories around the world \cite{R.Klockenkaemper2014}, GIXRF-based activities have been mostly confined to synchrotron radiation based instrumentation, which usually suffers from a decisive drawback: very limited accessibility. 

Even though no physical reason hinders a large-scale transfer of the technique to laboratory based instrumentation, there are not nearly as many examples of successful laboratory GIXRF instruments as one might expect. From a technical point of view, the requirements for a GIXRF laboratory setup differ only slightly from those for a TXRF or a X-ray reflectometry (XRR) instrument, both of which are available from multiple vendors and can be found in laboratories. For successfully creating an X-ray standing wave, one must have a monochromatic and low-divergence excitation source available just as for TXRF and XRR.

Thus, the reasons for the limited availability of laboratory GIXRF are to be found on the data evaluation side. First, profound knowledge of the instrumental function is required for effective GIXRF applications - especially knowledge of the effective solid angle of detection, which results from the interplay between the fluorescence detectors field of view and the incident angle-dependent projection of the spatial incident beam intensity distribution. Second, in almost all applications of GIXRF, a modeling of the experimental data is required in order to derive the information of interest. To create such a model, a problem-specific modeling approach must be developed, which is not always straightforward. As a result, GIXRF itself requires a lot of experienced input and thus could be considered less user-friendly than TXRF or XRR.

Several software packages have been developed in an effort to make the modeling aspect of GIXRF more user friendly \cite{F.Brigidi2017,D.Ingerle2016,M.K.Tiwari2016}. However, this addresses only part of the problem, as the effective solid-angle effects are usually either also part of the modeling parameter set in these software packages or they are not generalized enough so that they must be determined each time. This approach allows effective depth profiling results to be derived, as shown in several works (e.g. \cite{S.Motellier2013,D.Ingerle2014a,Maderitsch_2018,Pessoa_2018,Szwedowski_Rammert_2019}), but introduces unnecessary degrees of freedom that can degrade the reliability of the results obtained. This is especially true if parameters as the solid angle of detection are only taken into account using a qualitative model rather than using a full geometrical model as shown here.

This can be circumvented by taking a different approach: As the effective solid angle is mainly determined by the aperture geometry, by the distance of the detector from the sample surface as well as by the spatial intensity profile of the excitation beam, it can be determined in a dedicated approach for a given tool until any of the relevant parameters change. Furthermore, if an adequate geometrical model is used, even changes in relevant parameters of the setup can be taken into account and the new effective solid angle of detection can be predicted. 

In this article, we will demonstrate such an approach for the characterization of the geometry of a Bruker S4 T-Star. The approach is based on a well-known sample, which is pre-characterized using the reference-free GIXRF approach of PTB \cite{Hoenicke2019,M.Mueller2014}, for which the GIXRF signals of the laboratory tool can then be predicted. In addition to a characterization of the effective solid angle, this approach allows deriving information about the incident photon flux and other parameters that influence the overall detection efficiency. Thus, it paves the way for a full calibration of the GIXRF setup in order to enable quantitative results. Furthermore, this approach can be transferred to other types of GIXRF instrumentation, as it does not require any of the features specific for the Bruker S4 T-Star, e.g. the Mo-K$\alpha$ excitation.

\section{Experiments}
\subsection{Reference-free GIXRF-XRR at PTB}
At the PTB laboratory\cite{B.Beckhoff2009c} at the BESSY II electron storage ring, several beamlines and sets of intrumentation are operated that are dedicated to the radiometric calibration of both radiation sources and detectors. Apart from the existing infrastructure for a physically traceable calibration of X-ray detectors (e.g. photodiodes \cite{A.Gottwald2006,Gottwald_2010} and energy-dispersive detectors \cite{F.Scholze2009}), PTB offers unique possibilities for conducting X-ray fluorescence spectrometry \cite{Beckhoff2008}.

By relying on these radiometrically calibrated detectors and on reliable knowledge of the atomic fundamental parameters, well-characterized beamlines and a well-calibrated geometry of the instrumental setup, one can perform so-called reference-free X-ray fluorescence spectrometry \cite{B.Beckhoff2007a,Beckhoff2008}. Here, no certified reference material or calibration standards are needed in order to perform a fully traceable (traceable to SI units) quantitative analysis of the mass deposition of a given element. For these purposes, PTB operates several in-house developed instrumental setups \cite{B.Beckhoff2007a,J.Lubeck2013}, which can be installed at multiple beamlines at the BESSY II synchrotron radiation source. 

By performing the reference-free XRF analysis in different excitation conditions, especially with respect to the incident angle of the exciting radiation, depth-dependent information can be derived \cite{P.Hoenicke2009}. In the so-called grazing-incidence XRF regime (GIXRF), the X-ray standing wave (XSW) field arising from the interference of incident and reflected X-rays on a flat surface or interface results in incident angle-dependent changes to the excitation conditions. In combination with a modeling of the experimental data, this allows to derive depth-resolved information if the angular divergence of the incident beam is low enough. 

Such a modeling process can be performed on the basis of the Sherman equation \cite{Sherman1955}, which is shown below for application in GIXRF \cite{Hoenicke2019}. Here, the experimentally derived count rate $F(\theta_i,E_i)$ of a given fluorescence line of the element of interest, excited using photons of energy $E_i$ at an incident angle $\theta_i$ is the essential measurand. This count rate needs to be normalized on the effective solid angle of detection $\frac{\Omega(\theta_i)}{4\pi}$, the incident photon flux $\Phi_0$ and the detection efficiency of the used fluorescence detector $\epsilon_{E_f}$. By calculating the x-ray standing wave field intensity distribution $I_{XSW}(\theta_i,z)$, a numerical integration in conjunction with the depth distribution $P(z)$ of the element of interest and an attenuation correction factor, the experimental data can be reproduced. For a quantitative modeling, the atomic fundamental parameters, namely the photo ionization cross section $\tau(E_i)$ and the fluorescence yield $\omega_k$, and material-dependent parameters, e.g. the weight fraction $W_i$ of element $i$ within the matrix as well as the density $\rho$ of the matrix must be considered.

\begin{align}
\frac{4\pi\sin\theta_i}{\Omega(\theta_i)}\frac{F(\theta_i,E_i)}{\Phi_0\epsilon_{E_f}} &= W_i \rho \tau(E_i) \omega_k dz \cdot \sum_{z} P(z) \cdot I_{XSW}(\theta_i,z) \cdot \exp\left[-\rho\mu_{E_f}z\right]\textrm{.}
\label{eq:sherman}
\end{align}

It should be noted that the calculation of the XSW involves a full description of the sample. If concentration gradients are present, this must be taken into account by using appropriate depth-dependent optical constants. The in-depth composition of the sample is then changed in order to reproduce the experimental data by means of adequate optimization algorithms. A combined modeling of both the reference-free GIXRF and XRR data \cite{Hoenicke2019} or normal GIXRF-XRR\cite{M.Kraemer2006,Kraemer2007} is also possible. 

For such a modeling of experimental GIXRF data, profound knowledge of the relevant instrumental parameters is crucial. As already mentioned, the effective solid angle of detection $\frac{\Omega(\theta_i)}{4\pi}$ is very important as it is also strongly incident angle-dependent due to the changing projection of the incident beam profile onto the sample surface. Thus, even if no reference-free GIXRF is to be performed and thus parameters as the incident photon flux or the SDD's detection efficiency are not needed, the effective solid angle of detection must be known in order to be able to derive depth-dependent information from the experimental data. 

\subsection{Bruker S4 T-STAR for GIXRF}
The S4 T-STAR is a table-top TXRF spectrometer designed primarily for routine analysis of a wide variety of sample types. The instrument is equipped with a robotic sample changer that can handle a total of 90 sample carriers via 10 tray inserts.  In addition to widely used 30 mm quartz discs, other sample geometries such as microscope carriers and 2" wafers can be used. 

The instrument comprises two microfocus x-ray tubes (both powered with up to 50 W), one with a tungsten target and one with a molybdenum target. Three individually designed monochromators can be combined with these tubes. Each one is optimized for the characteristic fluorescence lines of the anode materials: Mo K$\alpha$ (17.4 keV) and W L$\alpha$ (8.4 keV), plus one for Bremsstrahlung (typically of the W anode) at 35 keV. All monochromators consist of a multilayer as the dispersive element, an entrance slit, a central aperture above the multilayer and an exit slit. The distance from the X-ray source to the center position of the multilayer is 85 mm and the distance from the center position to the sample is 100 mm. The monochromator for the GIXRF measurements with Mo K$\alpha$ was designed for the lowest possible beam divergence. A spherically polished (R: 14.5 m) Si crystal was used as a substrate and the applied Mo/Si multilayer coating has a variable spacing along the beam direction. The effective beam cross section at the height of the sample is 0.2 mm by 6.0 mm. Since the beam divergence is determined by numerous factors such as source size, position of apertures and sample as well as the properties of the multilayer, the actual beam divergence was determined experimentally. For this purpose, the beam profile between the exit slit of the monochromator and the sample position was scanned stepwise with an X-ray camera and then reconstructed in three dimensions. The beam divergence determined by this method is 0.014° \cite{Szwedowski_Rammert_2019}. The spectral bandwidth of the monochromator is relatively high due to the relatively compact design and was determined by monochromating the bremsstrahlung background to 700 eV. However, since the monochromator is adjusted to the Mo-K line, the effective bandwidth is significantly smaller and is in the range of the distance of the K$\alpha1,2$ lines plus the natural line width at about 120 eV.

The positioning of the X-ray tubes, the monochromators, the sample position and the excitation angle is done electromotively. One of the special features of the system is that it can optimize the positioning of all X-ray components in an automated measuring process. For this purpose, a sample stored in the device is loaded and a cyclic adjustment process of all relevant assemblies is carried out. This quality process can be activated automatically at regular intervals or as required and ensures reproducible measuring conditions even under changed ambient conditions and after the excitation conditions have changed. 

The sample is positioned by pressing it against three mechanical reference points in front of the silicon drift detector (SDD). A rotation and linear unit enable the plane spanned by the pressure points to be moved and tilted in space. Such a movement relative to the excitation beam can be used e.g. for a GIXRF measurement. The minimum step size for adjusting the incident angle is one motor step, which corresponds to about 0.0005$^\circ$.

\begin{figure}[!h]
  \centering
    \includegraphics[width=7cm]{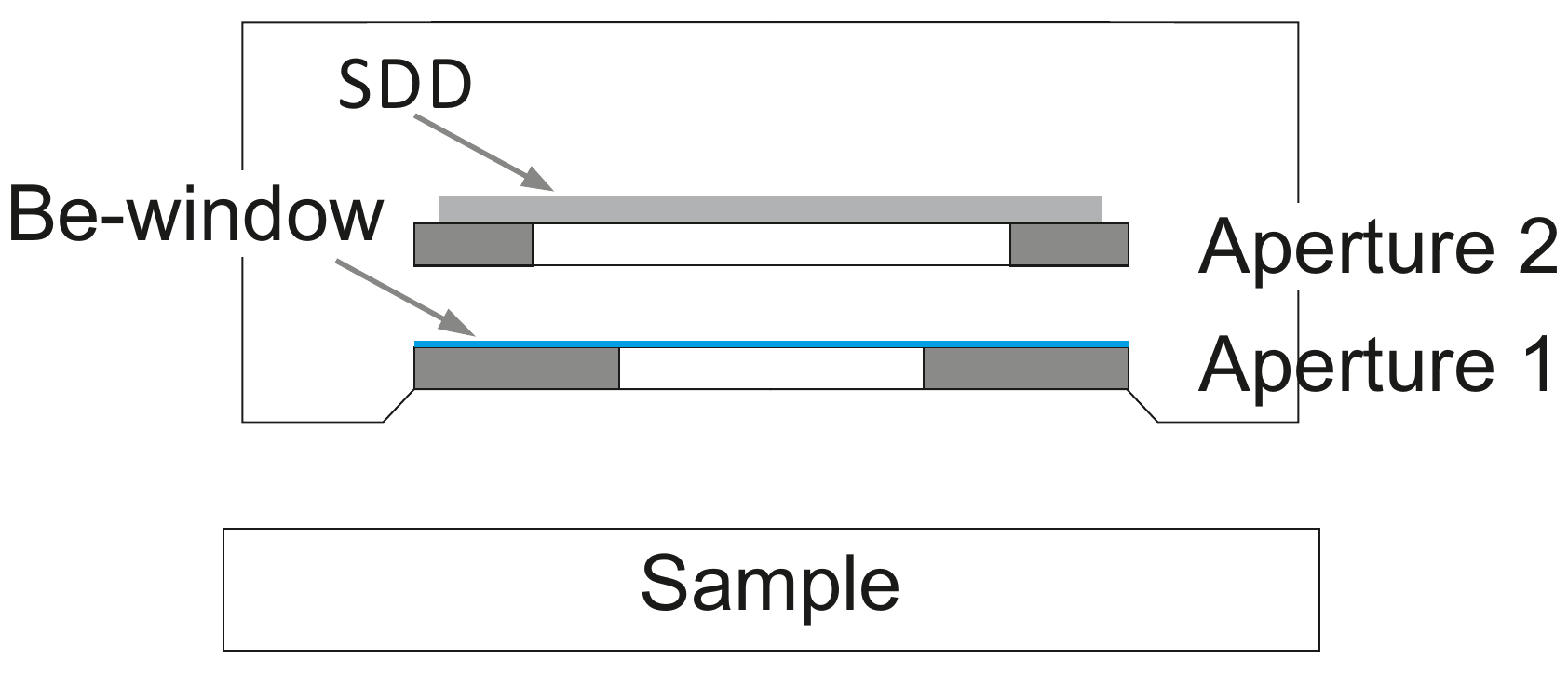}
  \caption{Sketch of the detector geometry inside the S4 T-STAR.}
  \label{fig:fig1}
\end{figure}

Based on the knowledge of the detector geometry (see fig. \ref{fig:fig1}) within the Bruker S4 T-STAR instrument, one can set up a geometrical calculation of the effective solid angle of detection. In addition to the exact sizes of the apertures, their distances and the sample detector distance, the lateral dimensions as well as the profile of the incident beam projection onto the sample surface are considered. The geometrical calculation is realized by first calculating the projection of the incident beam profile, which can be measured on the tool by scanning the sample and detector unit vertically through the beam, onto the sample surface depending on the incident angle. The solid angle of detection from each coordinate within the illuminated sample surface area is then calculated based on a geometrical consideration of the effective chip area and the effective distance between the chip and this coordinate, similar to the process in ref. \cite{B.Beckhoff2007a}. However, due to the rather large beam profile in the direction orthogonal to the plane of incidence and the very low detector distance to the sample surface, we here consider the full 2D illuminated area. Depending on the local relative incident intensity, the effective solid angle of detection is then numerically integrated in 2D in order to calculate the overall effective solid angle of detection for this incident angle.

\subsection{Characterization of Ni sample}
For a characterization of the geometrical parameters of the laboratory instrument, we used a Ni single layer deposited on a Si wafer piece by means of ion beam sputtering (IBS) at AXO DRESDEN GmbH. The nominal Ni layer thickness was 12 nm. However, oxidation in ambient air leads to the rapid formation of a thin oxide layer in the surface region of the sample. XRR measurements of the sample with Cu K-alpha radiation and fits of the XRR curves suggested a stack composition of ~11 nm Ni covered with ~2 nm NiO on the Si substrate (which also has a natural native SiO$_2$ top region). This sample was then characterized using the reference-free GIXRF-XRR technique of PTB in order to accurately model the in-depth structure to e.g. derive the oxide layer thickness. Therefore, we first performed reference-free GIXRF-XRR experiments at the four crystal monochromator (FCM) beamline \cite{Krumrey1998} of PTB at the BESSY II electron storage ring. An excitation photon energy of 9.0 keV was used in order to perform the experiments.

For the reference-free GIXRF-XRR experiments, PTB's in-house built instrumentation \cite{J.Lubeck2013} was used. The setup is installed in an ultra-high vacuum (UHV) chamber equipped with a 9-axes manipulator, allowing for a very precise sample alignment with respect to all relevant degrees of freedom. The emitted fluorescence radiation is detected by means of a calibrated \cite{F.Scholze2001a,F.Scholze2009} SDD mounted at 90$^\circ$ with respect to the incident beam. Calibrated  photodiodes\cite{B.Beckhoff2009c} on a separate 2$\theta$ axis allow for XRR measurements simultaneously with the reference-free GIXRF measurements as well as for the determination of the incident photon flux.

\begin{figure}[!h]
  \centering
    \includegraphics[width=16cm]{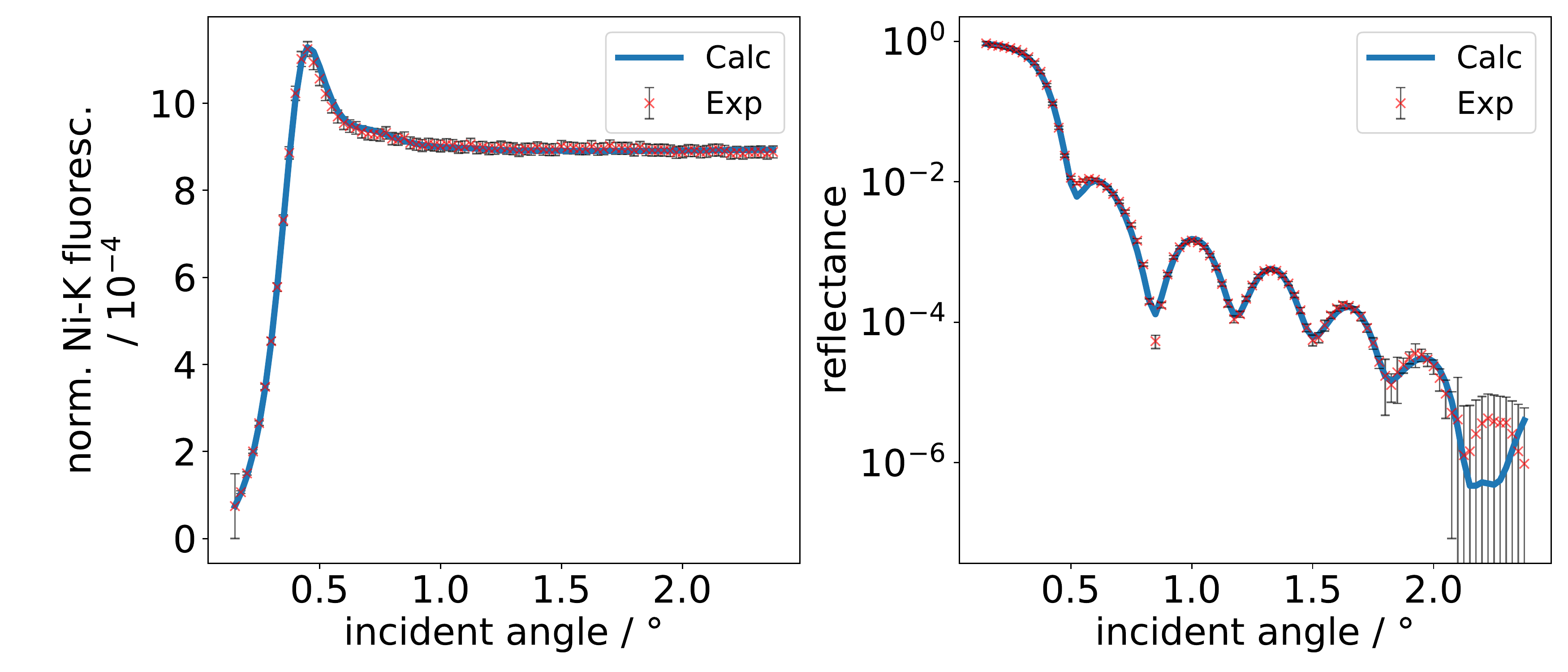}
  \caption{Comparison of the experimental GIXRF and XRR curves measured at PTB using an excitation photon energy of 9.0 keV and the corresponding model calculations (in red).}
  \label{fig:fig2}
\end{figure}

At each incident angle, the recorded fluorescence spectra are deconvolved using detector response functions \cite{F.Scholze2009} for the relevant fluorescence lines as well as for the background contributions. The experimental data is normalized with respect to the sine of the incident angle, the incident photon flux, the effective solid angle of detection as well as the integration time as shown in eq. \ref{eq:sherman}. The experimental data sets for the Ni GIXRF and the XRR derived in this way are shown in fig. \ref{fig:fig2} in black, including error bars. The uncertainties consist of a contribution from the solid-angle calculation and from the counting statistics for GIXRF and of a footprint correction contribution and a dark-current contribution for the XRR data.

The modeling of the reference-free GIXRF-XRR\cite{Hoenicke2019} assumes a layer structure consisting of a surface NiO layer, a metallic Ni layer and a Si substrate which is covered with a thin native oxide layer. For each layer, concentration depth profiles are calculated that are allowed to intermix at the interfaces. These concentration depth profiles are then used to calculate depth profiles for each optical constant ($\delta$ and $\beta$). Here, bulk optical constants for NiO, Ni, SiO$_2$ and Si from \cite{T.Schoonjans2011} were used and scaled with the modeled material densities. In the intermixed regions, the effective optical constants were calculated accordingly by means of a linear combination.

The full layer stack is separated into sublayers in order to calculate both the resulting XRR curve and the XSW using an implementation of the matrix algorithm (XSWini\cite{Pollakowski}). The derived intensity distribution within the XSW is then numerically integrated in conjunction with the calculated concentration depth profiles and all other relevant instrumental and fundamental parameters to calculate the angular fluorescence profile for Ni in a quantitative manner as shown in reference \cite{P.Hoenicke2009}. We have employed an experimentally determined fluorescence yield for the Ni-K shell \cite{Menesguen2017, Guerra2018} and tabulated fundamental parameter data from \cite{T.Schoonjans2011}. For a modeling of the incident beam divergence, multiple different incident angular responses are averaged. Using this technique, a layer stack consisting of  a 1 nm nickel oxide layer on a 10 nm Ni layer was found. The native oxide layer on the Si substrate has a thickness of about 1.4 nm.

\section{Results and Discussion}
\subsection{Determination of effective solid angle of detection}
A GIXRF measurement on the pre-characterized Ni sample was also performed on the Bruker S4 T-STAR tool. The recorded fluorescence spectra were treated in a similar way as for the reference-free GIXRF experiments at PTB in order to derive the Ni-K as well as the Si-K GIXRF curves (sum of the K$\alpha$ and K$\beta$ lines). The laboratory data set can only be normalized with respect to the sine of the incident angle and the integration time, as both the incident photon flux and the effective solid angle of detection are not known absolutely.

For a determination of these parameters, we use knowledge about the previously characterized Ni sample. Using the verified layer model, we can now calculate both the Ni-K and the Si-K fluorescence intensities the sample would emit for a Mo-K$\alpha$ excitation using eq. \ref{eq:sherman} (right-hand side). As this is performed in a quantitative manner, the experimental data should exactly match the two datasets if normalized to the correct incident photon flux, correct detector efficiency and correct effective solid angle of detection (left-hand side of eq. \ref{eq:sherman}). This can then be used to define a criterion for an optimization algorithm that uses the geometrical parameters of the detector geometry and an absolute incident photon flux as parameters. As the Bruker S4 T-STAR setup is working in air, also the attenuation in air from the sample to the entrance window of the detector must be taken into account.

The optimization algorithm varies the relevant geometrical parameters e.g. the aperture sizes and distances and the Be entrance window thickness within the fabrication tolerances and calculates the effective solid angle of detection for each incident angle in the experimental data. The shape of the incident beam intensity profile has been derived from a dedicated experiment during the sample alignment procedure and is only varied with respect to its widths and position relative to the field of view of the detector. In addition, the incident photon flux is varied by means of the optimization routine in order to match both the angular shape (sensitive for the solid angle of detection) and the absolute value of the correctly normalized experimental data (sensitive to the incident flux, the Be-window thickness and the solid angle of detection). Similarly as for the pre-characterization of the Ni sample, the angular divergence of the incident beam is taken into account by averaging over multiple different incident angles in the vicinity of the angle of interest.

\begin{figure}[!h]
  \centering
    \includegraphics[width=14cm]{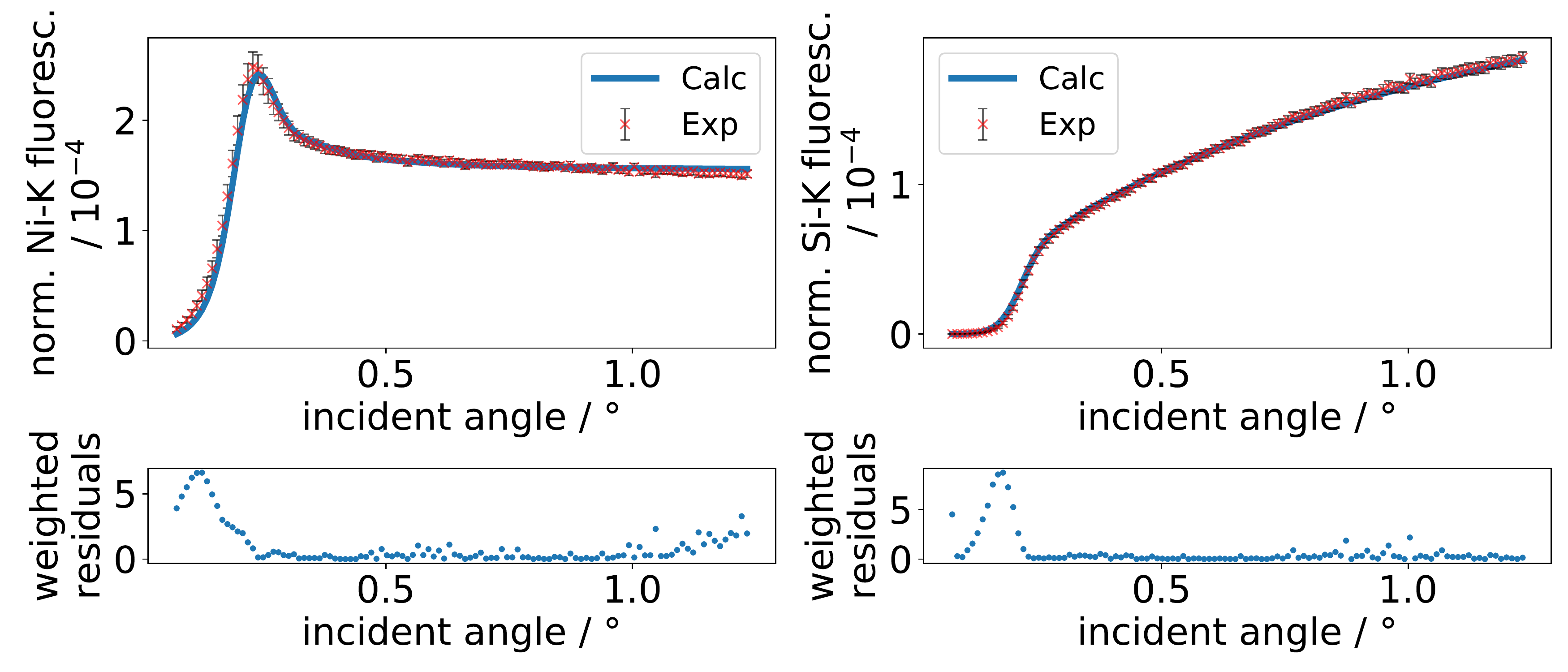}
  \caption{Comparison of the calculated fluorescence signals for Ni-K (left side, solid line) and Si-K (right side, solid line) to the experimental data (dotted lines) normalized to the effective solid angle of detection and photon flux. The plotted error bars of the experimental data were calculated using the counting statistics as well as an approximation of the uncertainty of the effective solid angle of detection, which increases towards lower incident angles.}
  \label{fig:fig3}
\end{figure}

In fig. \ref{fig:fig3}, the result of this optimization is shown for the calculated Ni (left side, solid line) and Si (right side, solid line) fluorescence signals in comparison to the experimental data (dotted lines) normalized to the effective solid angle and photon flux (for this experimental run, a photon flux of $4.3 \times 10^7 s^{-1}$ was determined). The remaining residuals, which are calculated by dividing the square of the local difference between calculated and experimental signals by the square of the experimental uncertainty (they consist of a contribution from the solid-angle calculation and from the counting statistics), are also shown and indicate remaining discrepancies at very low incident angles as well as at high incident angles for the Ni data. The latter are caused by a relatively large spectral background in the SDD spectra resulting in slightly underestimated Ni signals. The low incident angle deviations are caused by artifacts, which are due to a rather difficult sample mounting. Due to the fixed diameter of the circle on which the mechanical reference points in front of the detector are mounted, we had to put the smaller Ni-layer sample in a specially designed tray. This can result in a small height offset between the sample surface and the tray surface and, thus, either shadowing effects or direct excitation of the sample edge at low incident angles. If larger samples are used, these artifacts can be reduced.

\begin{figure}[!h]
  \centering
    \includegraphics[width=7cm]{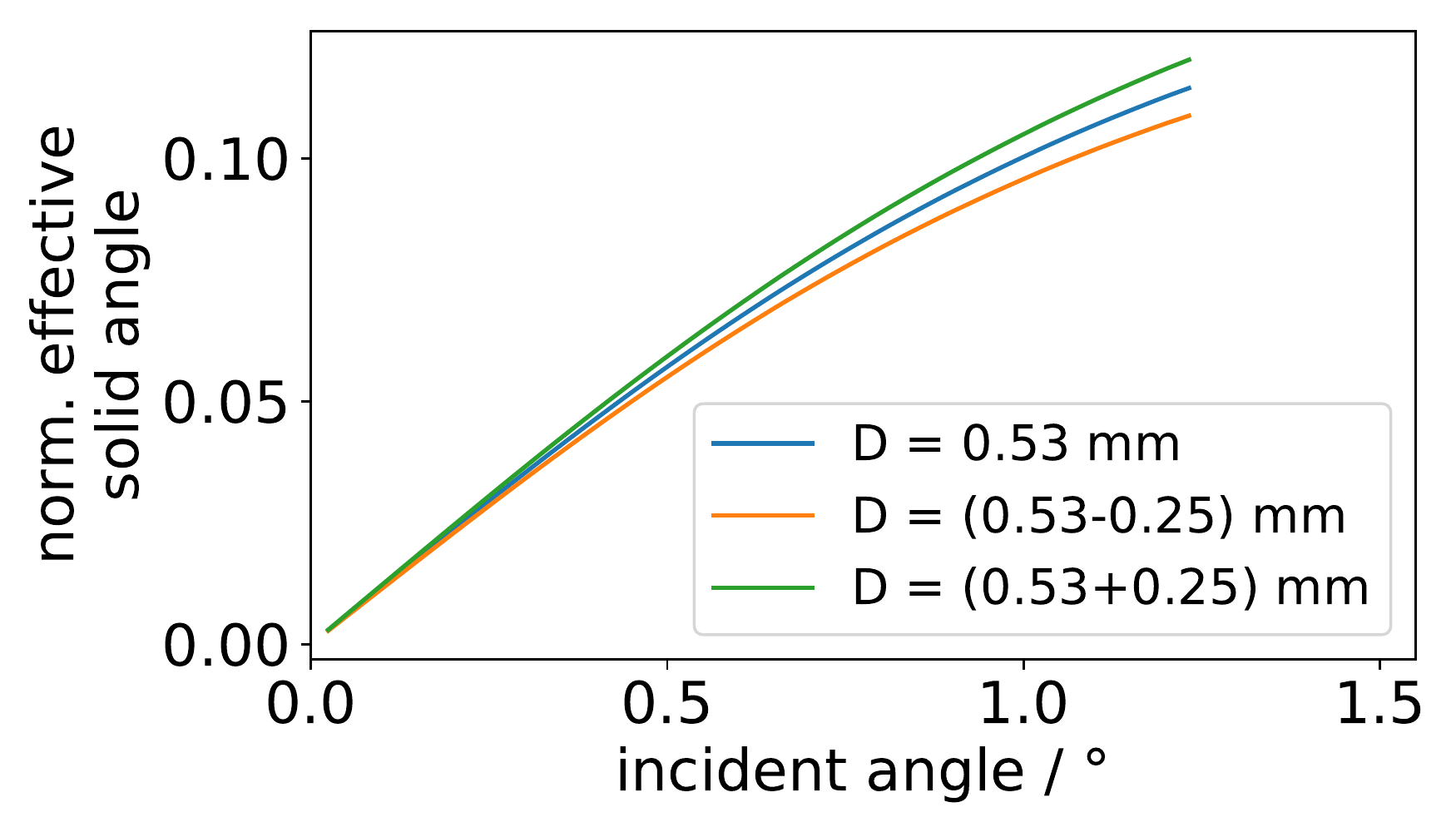}
  \caption{Comparison of the effective solid angle calculated for the detector distance as determined (blue curve) as well as for two other detector distances.}
  \label{fig:fig4}
\end{figure}

Using this methodology, the geometrical parameter set that defines the effective solid angle is determined. As the effective solid angle is calculated using a geometrical model based on these parameters, it is now also possible to change parameters of the setup, e.g. to change the sample detector distance (defined as the distance between the sample surface and the surface of apperture 1 in fig. \ref{fig:fig1}) without having to repeat the characterization procedure. The changed parameters can be adopted and the effective solid angle can be re-calculated for the new geometry as shown in fig. \ref{fig:fig4}. The shown effective solid angle for a detector distance of 0.53 mm is the one derived from the previously presented characterization. 

\subsection{GIXRF characterization of ion implant samples}
For a validation of the parameter set derived for the calculation of the instrument's effective solid angle of detection function, two ultrashallow ion implant samples were used. Both arsenic implants were put into silicon. Sample A was implanted at 1.25 keV and a total dose of $5 \times 10^{14} \frac{At.}{cm^2}$, whereas sample B was implanted at 2.5 keV and a total dose of $1 \times 10^{15} \frac{At.}{cm^2}$.

\begin{figure}[!h]
  \centering
    \includegraphics[width=14cm]{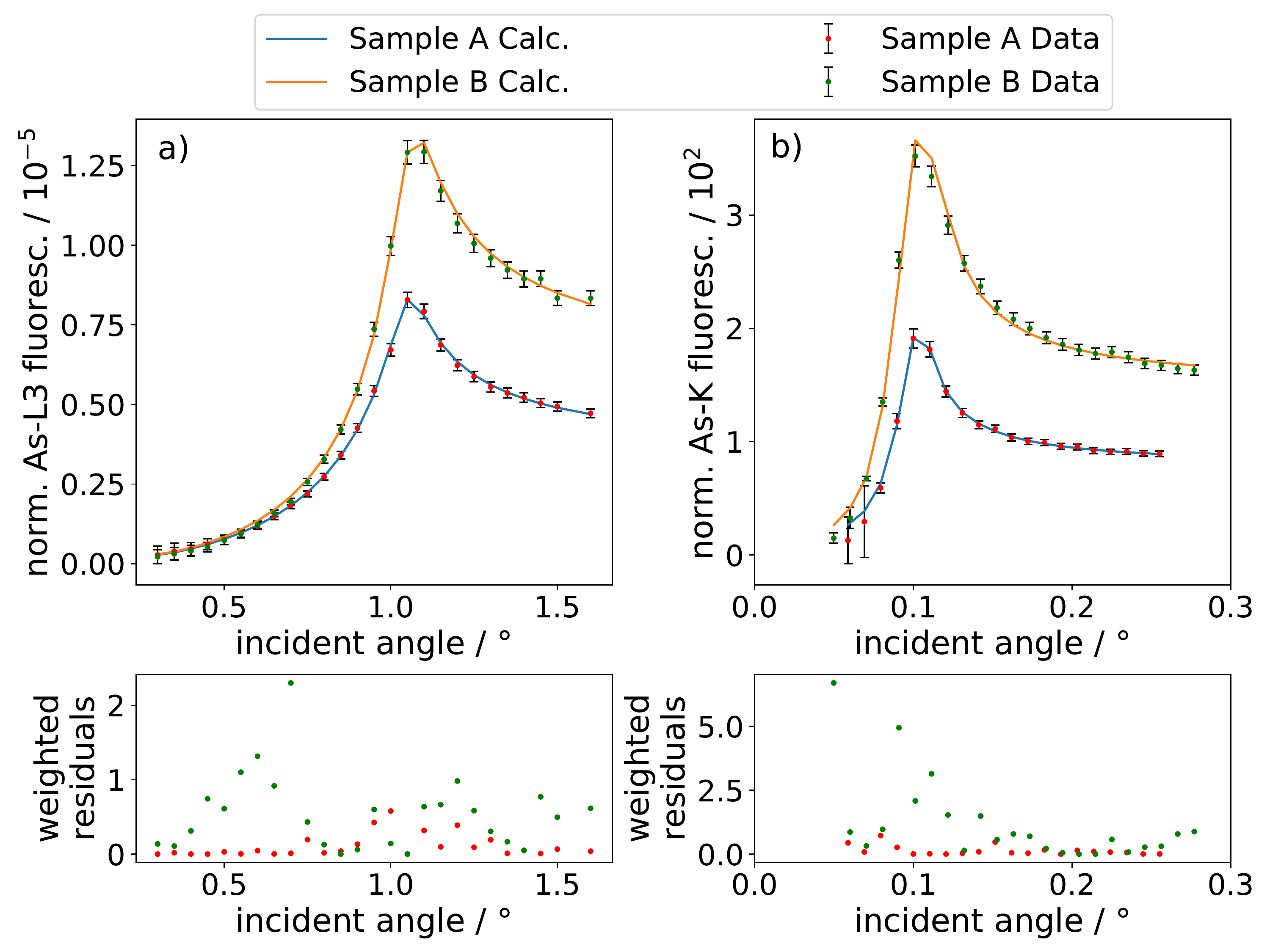}
\caption{Comparison of the experimental GIXRF and the calculated curves for both implant samples and the experiments performed at PTB (plot a) and at Bruker with the Bruker S4 T-STAR (plot b). The plotted error bars of the experimental data were calculated using the counting statistics as well as an approximation of the uncertainty of the effective solid angle of detection, which increases towards lower incident angles.}
  \label{fig:fig5}
\end{figure}

These samples were pre-characterized using PTB's reference-free GIXRF approach, and depth profiles for the implant depth distribution were derived according to \cite{P.Hoenicke2009}. These GIXRF experiments were conducted at the plane grating monochromator (PGM) beamline \cite{F.Senf1998} of PTB using an incident photon energy of 1540 eV. The same instrumentation end experimental procedure as for the Ni sample was applied and the normalized fluorescence signal derived for the As-L3 lineset (sum of all relevant As-L3 fluorescence lines) was used for the depth profiling. For the modeling, the XSW calculation was performed assuming an undisturbed silicon wafer (optical constants were taken from \cite{T.Schoonjans2011}) and an asymmetric Gaussian distribution was used for the As implantation depth profile \cite{P.Hoenicke2009}. In order to derive the implanted depth and the shape parameters from a given implant energy, parameter relations were derived from TRIM calculations \cite{J.F.Ziegler2010} and used to reduce the degrees of freedom for the modeling. In fig. \ref{fig:fig3} (left side), the obtained experimental GIXRF data as well as the corresponding modeled curves are shown for the two samples for the experiments performed at PTB. Total retained As doses of $6 \times 10^{14} \frac{At.}{cm^2}$ for sample A and $1.1 \times 10^{15} \frac{At.}{cm^2}$ for sample B were quantified from the depth profiles. These values are in line with the nominal data even though the implanted depths determined are somewhat higher than expected (see fig. \ref{fig:fig5} for the derived depth profiles).  

Due to relatively high excitation photon energy (Mo-K$\alpha$) in the Bruker S4 T-STAR, the depth profiling performance for the identical two samples crucially depends on both the correct calculation of the effective solid angle and the correct alignment of the sample. Furthermore, the quantification of the total retained As doses crucially depends on the accuracy of the previously determined photon flux, Be window thickness and the effective solid angle of detection. Thus, a comparison of depth profiling results as derived from the experiments on the Bruker S4 T-STAR allows for an assessment of the quality of the previously derived instrumental parameters as any errors would result in either different implant doses or different depth distributions. 

\begin{figure}[!h]
  \centering
    \includegraphics[width=8cm]{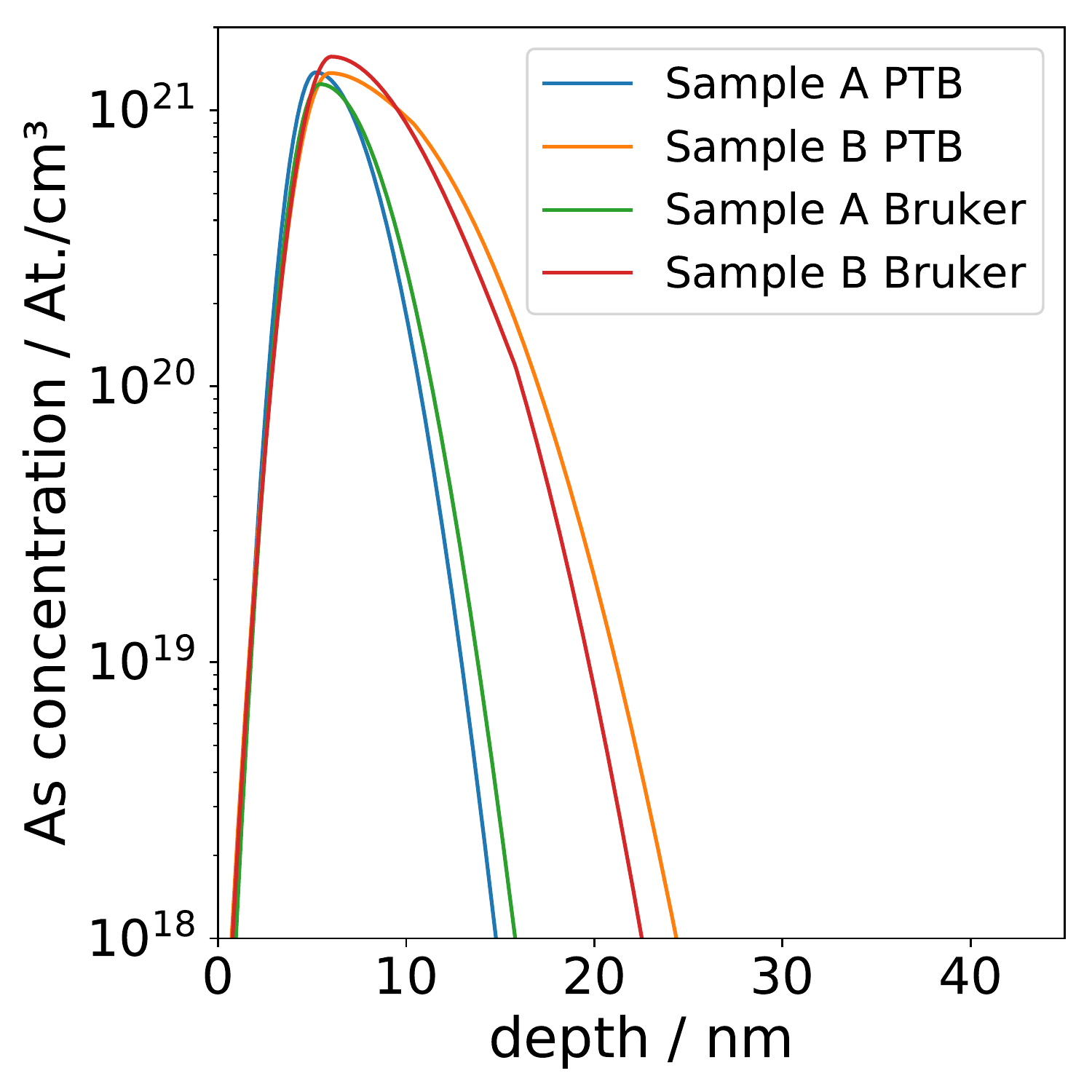}
\caption{Comparison of the As depth profiles derived from the modeling of the GIXRF experimental data obtained at PTB and with the Bruker S4 T-STAR instrument.}
  \label{fig:fig6}
\end{figure}

In the applied experimental procedure, the GIXRF experiment on the Ni sample was first run in order to derive the effective solid angle and the incident photon flux. For this experimental run, a photon flux of $4.3 \times 10^7 s^{-1}$ was determined as already mentioned. Afterwards, the two implant samples were measured after proper alignment. Count rates for the As K-shell lines (sum of the K$\alpha$ and K$\beta$ lines) were derived from the spectra at each incident angle, normalized to the effective solid angle and then modeled using the same approach as for the experiments performed at PTB. The experimental GIXRF data obtained as well as the corresponding modeled curves are shown in fig. \ref{fig:fig5} (right side). The derived As depth profiles are shown in fig. \ref{fig:fig6} in comparison to those derived from the experiments using the PTB setup. Both the agreement between the experimental and modeled GIXRF curves and between the two sets of derived depth profiles is very good. For sample B, the Bruker-derived depth profile is slightly shifted towards the surface, whereas the profile of sample A is shifted away from the surface. The quantified total retained As doses of $5.9 \times 10^{14} \frac{At.}{cm^2}$ for sample A and $1.1 \times 10^{15} \frac{At.}{cm^2}$ for sample B are however agreeing very well with the PTB results. Due to the much higher excitation photon energy, the modeling of the data measured on the Bruker S4 T-Star is much more sensitive to uncertainties in the incident angular scale. Thus, a marginally misaligned angular scale (= an angular offset) and the very low measurement point density is the origin of the shifted depth profiles. However, this can be improved by measuring at higher angular resolution especially in the vicinity of the critical angle. 

\section{Conclusion}
In this article, a methodology to characterize both the effective solid angle of detection and the incident photon flux of a Bruker S4 T-STAR instrument has been presented and validated by depth-profiling well-known As ion implant samples. By first characterizing the Bruker tool, we were able to derive quantitative results for both the in-depth distribution and the total retained doses, which are competitive with reference-free GIXRF results obtained using dedicated instrumentation on a synchrotron radiation beamline. 

Even though we only used shallow ion implant samples for validation and demonstration of the depth-profiling capabilities of the Bruker S4 T-STAR instrument in conjunction with the characterized effective solid-angle properties as well as the incident photon flux, it can be expected that the S4 T-Star will also work for other types of sample structures. This includes above-surface structures (e.g. nanoparticles and nanostructures) as well as subsurface structures (e.g. thin layer stacks or diffusion depth profiles). However, especially for the above-surface structures, the sample mounting must be improved in order to address the related modeling deviations at very low incident angles. The range of possible materials which could be studied with the instrument is limited on the one hand side by the employed excitation and fluorescence detector. Low-Z materials up to aluminum cannot be studied at present, as their fluorescence is nearly fully attenuated in the Be detector window and high spectral backgrounds due to silicon fluorescence is to be expected for higher incident angles. In addition, of course also the element-dependent detection limits of the tool must be considered.

By expanding the characterization procedure with additional materials other than Ni and Si, one may expect it to also provide information about the photon energy-dependent detection efficiency of the equipped fluorescence detector or the influence of secondary excitation due to photo-electrons. This would then enable a quantification of fluorescence lines, which are energetically further away from the elements used for the characterization and thus outside of the energy region where the detector efficiency is very close to unity. This will enable a full calibration of the laboratory-based GIXRF setup and, thus, allow a fully quantitative analysis without any further calibration by means of other standards or reference samples. In addition, the approach presented here is also transferable to other types of TXRF and GIXRF instrumentation from other vendors or even custom-built setups.

\section{Acknowledgements}
This research project was performed within the EMPIR projects Aeromet and Adlab-XMet. The financial support of the EMPIR program is gratefully acknowledged. It is jointly funded by the European Metrology Programme for Innovation and Research (EMPIR) and participating countries within the European Association of National Metrology Institutes (EURAMET) and the European Union.

\bibliographystyle{unsrt}  
\bibliography{references}  

\end{document}